# Controlling Silver Nanoparticle Size and Morphology with Photostimulated Synthesis

A. K. Popov[1,2,3], J. Brummer[2], R. S. Tanke[2], G. Taft[3], A. Wruck[2,4], M. Loth[3,4], R. Langlois[4,5] and R. Schmitz[5]

*University of Wisconsin-Stevens Point, Stevens Point, Wisconsin 54481*

[1]Corresponding author, [2]Department of Chemistry, [3]Department of Physics & Astronomy, [4]Undergraduate students, [5]Department of Biology

**Abstract**. Photo-induced synthesis and control over the size and shape of colloidal silver nanoparticles is investigated in contrast to photo-stimulated aggregation of small nanoparticles into large fractal-type structures. The feasibility of light-driven nanoengineering which enables manipulation of the sizes and shapes of the isolated nanoparticles is studied by varying the amount and type of the stabilizing agent and the type of optical irradiation.

## 1. Introduction.

One of the remarkable properties of nanoparticles is their high surface-to-volume ratio which leads to striking optical properties not available in bulk materials. In particular, metal nanoparticles acquire optical resonance, which depends not only on the free electron density, but also on the size and shape of the particles. This is in strong contrast to bulk metals where the optical properties are primarily determined by the plasma frequency, i.e., by the number density of free electrons. Such geometrical resonance attributed to metal nanoparticles is called "plasmonic" resonance. Thus, in optical processes like absorption, nanoparticles act like artificial atoms. Anisotropic nanoparticles possess several resonance frequencies, which depend on the polarization of light. Since the resonance frequencies depend on the size and shape of the particles, absorption spectra and nonlinear-optical properties of a material composed of a mixture of such nanoparticles depend on their distribution over different sizes and shapes.[1-5] This can be used to monitor changes in the nanoparticles' morphology and size during their synthesis.

Manipulating geometrical parameters of nanoparticles presents a nanoengineering problem of paramount importance. As shown by Shalaev, Stockman and co-workers[1], among the striking properties of metal nanoparticles is their ability to dramatically enhance the local optical electromagnetic near-fields. Concentration of electromagnetic radiation energy in subwavelength zones and, therefore, breaking the fundamental diffraction limits in optics enables numerous novel exciting applications in nanophotonics, for creation of ultrasmall lasers and opto-electronic devices, in biophotonics, for sensing of molecules, and for photomodification of biological cells.[4,6] Nanoparticles can also be fixed in various solid matrices and thus form metamaterials. Nanostructured silver metamaterials made of nanoparticles of a special shape recently attracted a great interest in the context of the proposal for creation of 'left-handed' materials with negative refractive index in the optical range.[7(a,b)] Recent successful experiments on realization of superlenses based on silver metamaterials[7(c)] are among the most exciting advances in nanoscience.

Light-driven nanoengineering can be applied to different synthetic methods which include photoinduced synthesis in colloids[1-6,8,9] and laser ablation.[10] The technique of photostimulation in colloids employs reduction of cations in solution. The basis of this technique is to take a metal ion and reduce it to a neutral metal species while in solution. The reduction takes place in the presence of a stabilizing agent, which is attracted to the metal's surface. The coating shields nanoparticles and thus prevents their further growth. This results in a colloidal solution of metal nanoparticles in contrast to a solution with a bulk metal precipitate. Light produces a size and shape-dependent photo-effect which may break the stabilization effect of the polymer shield and, thus, may trigger further nanoparticle growth or clusterization with other nanoparticles. A possible relationship between this photo-effect and the anisotropic plasmon resonance was investigated by Mirkin and co-workers.[5] Complex electrochemical processes involved in photoassisted synthesis of metal nanostructures in colloids[2,3,6,7] are determined by the properties of the stabilizing agents, the thickness of the stabilizing coating, and the electrolytic properties of the solution. In previous work, photostimulation led to growth of aggregates consisting of hundreds to thousands of small metal nanoparticles,[2-4,8,9] whereas in other cases a growth and change in the shapes of isolated nanoparticles were observed.[5] Detailed mechanisms of different light-driven nanoengineering methods remain to be determined. Since nanoparticle growth is triggered by the light source, it would be interesting to determine the extent of particle changes that occur with various light sources and stabilization methods. In this account we report further investigation of the synthetic method,[5] which predominantly leads to photoinduced changes of size and shape of silver nanoparticles instead of clustering of small nanoparticles into aggregates. The study is aimed at providing experimental data and gaining further insight into the specific features of such processes and shape-selective synthesis strategies. The monitoring of photoinduced modification of the nanoparticle shape was performed indirectly through the evolution of the colloid absorption spectra and confirmed directly by characterization with transmission electron microscope (TEM) images of the synthesized nanoparticles. Various nanoengineering options are demonstrated, which include variation of the particles sizes at the nanoscale as well as conversion of metal nanospheres to ellipsoids and nanoprisms. Formation of small nanoparticles and their aggregation into large multiparticle fractal-type structures will be reported elsewhere.

**2. Synthetic Technique**.

In our experiments, silver nanoparticles were produced in colloidal form by reduction of $AgNO_3$. Two methods which provided a different coating of nanoparticles were used. In one method, the reduction was performed by $NaBH_4$ with initial stabilization by citrate and Bis(p-sulfonatophenyl) phenylphosphine dihydrate dipotassium (BSPP) salt. Photostimulation by broadband visible lamps as well as by argon-ion ($\lambda$=514.5 nm) and helium-neon ($\lambda$=632.8 nm) lasers was investigated. In the second method, the reduction was accomplished with ethanol with initial stabilization by the polymer (poly)vinyl pyrrolidone (PVP). Photostimulation with a mercury-vapor lamp as well as with an argon-ion laser ($\lambda$=514.5 nm) was investigated.

## 3. Results and Discussion.

*3.1 Borohydride method.*

In the first (borohydride-based) method, an aqueous colloidal solution of silver nanoparticles was synthesized at room temperature by adding 1 mL of 50 mM sodium borohydride to 100 mL of a solution of 0.1 mM silver nitrite and 0.3 mM of trisodium citrate while stirring the solution. Then 2 mL of 5 mM BSPP was added slowly. After the solution was mixed for a few minutes, it was transferred into the appropriate number of cuvettes. Initial ultra-violet/visible (350-800 nm) absorption spectra were obtained at room temperature using a Varian Cary 300 Spectrophotometer. After the initial spectrum was obtained, irradiation was started and continued until the absorption spectrum no longer changed. Typical irradiation periods lasted several days for lamps and the He-Ne laser, while only several hours for the argon laser. Intermediate absorption spectra were obtained at various times during the course of the irradiation in order to monitor the changes in the colloid. A final absorption spectrum was also obtained for each irradiated solution and for a sample of the initial solution that was stored in the dark. Once all spectra were obtained, a microscopy slide was prepared for each solution, and TEM images were obtained.

Jin, et. al.[5] reported that synthesis of nanoprisms was initiated by light at wavelengths between 350 and 700 nm. The results of our experiments presented below show, however, that the photostimulation effects are different for the two synthetic methods we used and for different light sources falling within this spectral range. The strongest modification of the nanoparticle shapes was observed for the borohydride-based method with irradiation from an ordinary compact fluorescent bulb (Westinghouse Model Twist 40W fluorescent). The electrical power of this lamp was 40 W, and it produced an optical power density of 7 mW/cm$^2$ measured at the sample location. The broadband spectrum of this lamp is shown in **Figure 1** and covers the visible range (400-700 nm).

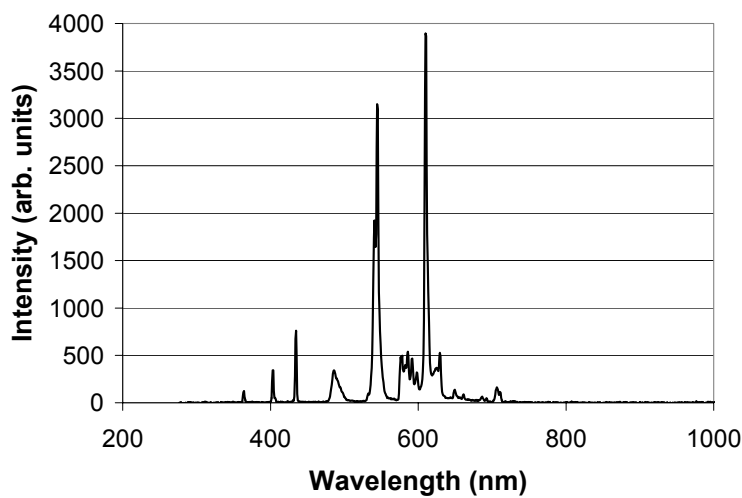

**Figure 1**. Radiation spectrum for the compact fluorescent bulb used in the experiments.

**Figures 2-4** show changes in the absorption spectrum and corresponding changes in the size and shape of the nanoparticles produced by this method. **Figure 2** shows drastic changes in the absorption spectrum in the course of irradiation. For ease of comparison each spectrum was first normalized to its maximum and then

shifted so that each spectrum has the same value at 350 nm. The most striking features concern the appearance of absorption for wavelengths greater than 450 nm, the red shift of the main absorption peak from 400 nm to 444 nm, and the appearance of additional absorption peaks at 500 nm and 650 nm. This is in good qualitative agreement with the spectral changes observed by Jin, et. Al.[5] No significant changes occurred for the sample kept in the dark, which indicates that the irradiation indeed affected the solution, and the growth observed was not simply due to thermal effects which occurred during the elapsed time. The main characteristic changes in the irradiated sample occurred within about 24 hours.

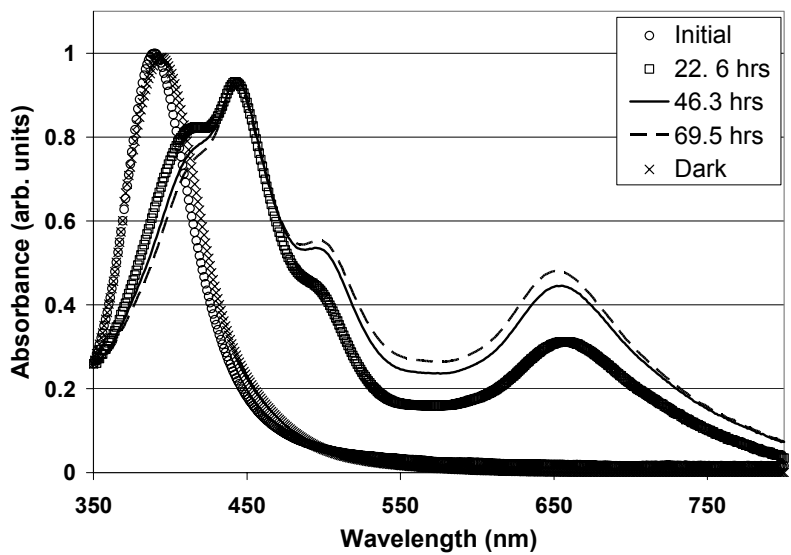

**Figure 2**. Evolution of the absorption spectrum of the silver colloid irradiated with a fluorescent lamp.

**Figures 3-5** present direct evidence of photostimulated changes in the size and shape of colloidal silver nanoparticles. **Figure 3** shows typical nanoparticles in the initial stabilized phase of the freshly prepared colloid before the irradiation begins. The characteristic nanoparticle sizes range from a few nanometers to about 20 nm. **The** low-resolution TEM image presented in **Figure 4** and the higher-resolution image shown in **Figure 5** confirms that photostimulation leads to the conversion of small spherical nanoparticles into large asymmetric and symmetric nanoparticles of different shapes, including symmetric triangular nanoprisms with a size of about 60 nm. The conversion is usually followed by the formation of some small aggregates. TEM images for the control sample kept in the dark look similar to **Figure 3**.

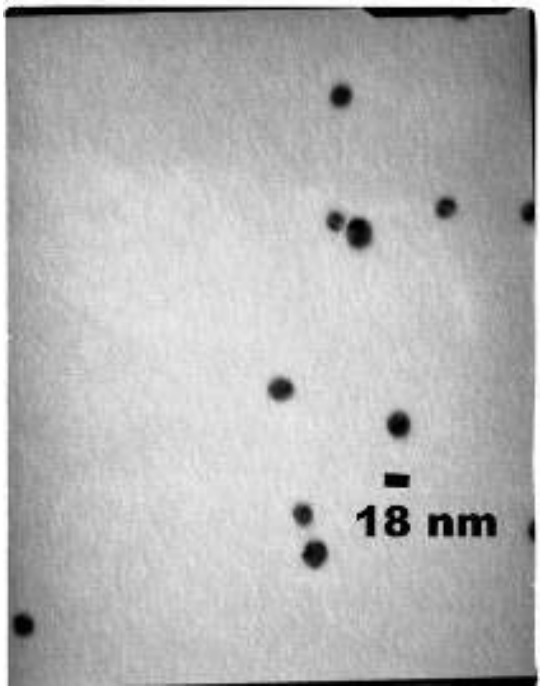

**Figure 3**. A typical TEM image of silver nanoparticles in the original colloidal solution produced by the reduction of $AgNO_3$ with $NaBH_4$.

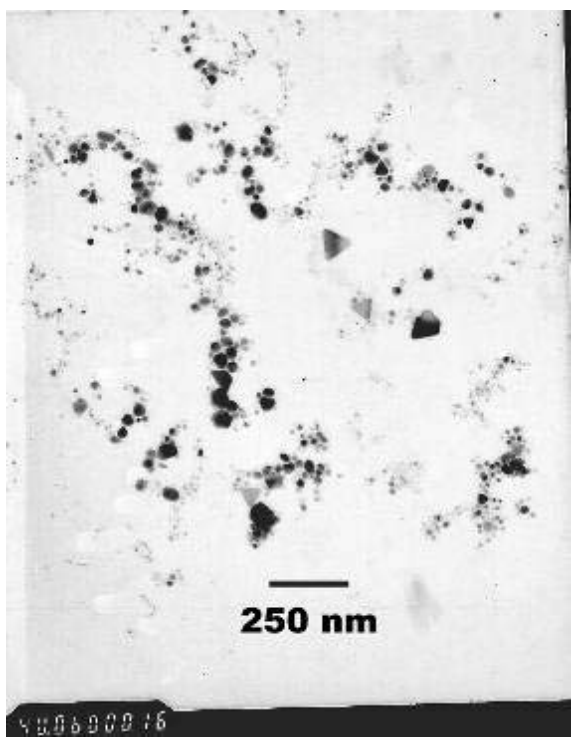

**Figure 4.** A mixture of silver nanoparticles of different sizes and shapes including nanoprisms and clusters of small nanoparticles that appeared after irradiation of a $AgNO_3/NaBH_4$ silver colloid by the fluorescent lamp.

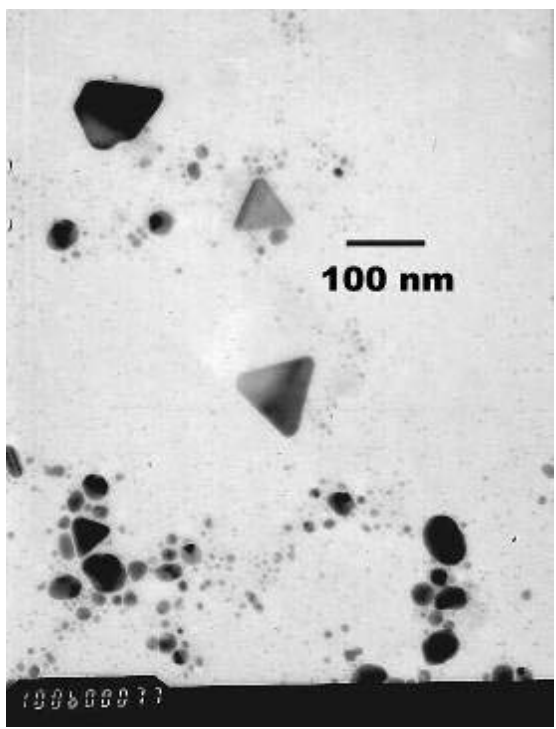

**Figure 5.** TEM image of silver nanoprisms and elliptical nanoparticles formed in a AgNO$_3$/NaBH$_4$ silver colloid under irradiation by the fluorescent lamp.

Irradiation with an argon-ion laser also caused dramatic changes in the colloid absorption spectrum as shown in **Figure 6**. The development of the secondary peak at a wavelength of 575 nm indicates an abundance of larger nanoparticles, which have a lower-frequency plasmonic resonance. The development of the long wavelength tail in the absorption spectrum indicates the formation of multi-particle aggregates and/or a variety of nanoparticles of even larger sizes. This absorption spectrum has fewer peaks as compared to the absorption spectrum of the sample irradiated with the broadbandwidth fluorescent bulb. This could result from the excitation of fewer plasmonic resonances by the monochromatic argon-ion laser beam.

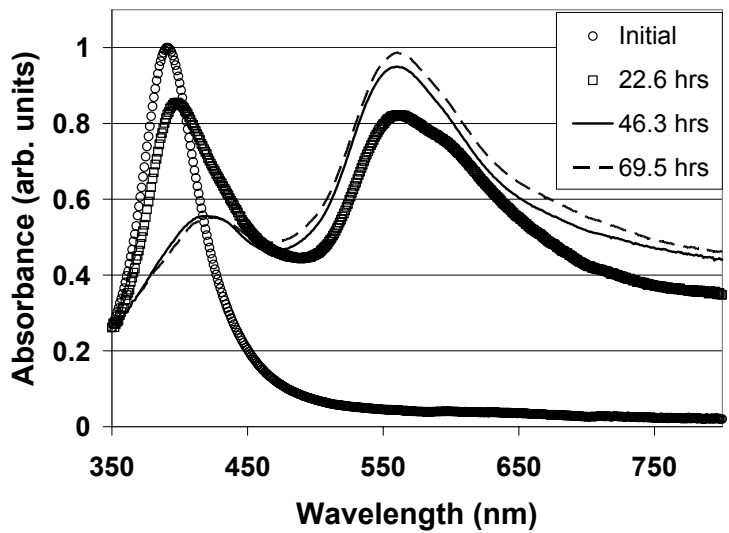

**Figure 6.** Absorption spectrum for the colloid irradiated by an argon-ion laser.

Photo-stimulation effects of UV light was also investigated by using a BLE-1800B Spectronic bulb (cylindrical, 18" long) in a Spectroline Model X-15A housing (120V, 60 Hz, 0.35 amps). The lamp produced an irradiance of 1.7 mW/cm$^2$ at 10 cm from the bulb, where the colloid was placed. As shown in **Figure 7**, the main part of the spectrum is centered at 365nm. **Figure 8** shows typical changes in the absorption spectrum of the colloid caused by irradiation with this UV lamp. Although the spectrum changes are less significant than in the previous case, the shift in the peak suggests a slight increase in the average size of the nanoparticles. As shown in **Figure 9,** the irradiation with a He-Ne laser caused even less of a change.

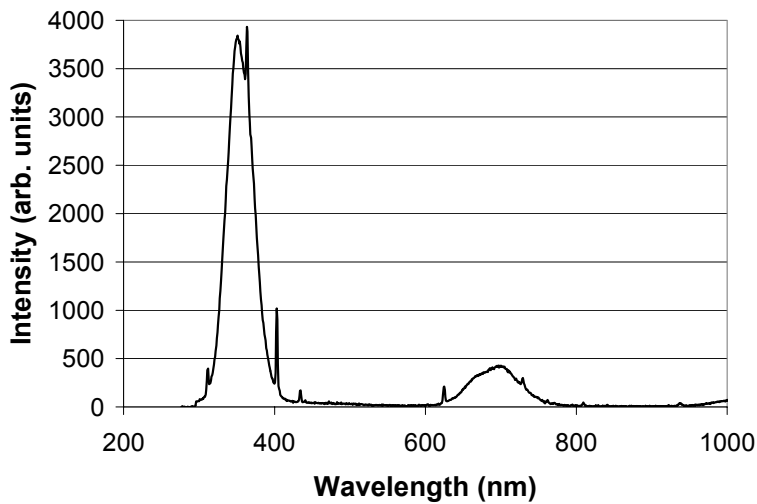

**Figure 7**. Radiation spectrum of the UV lamp used for photostimulation.

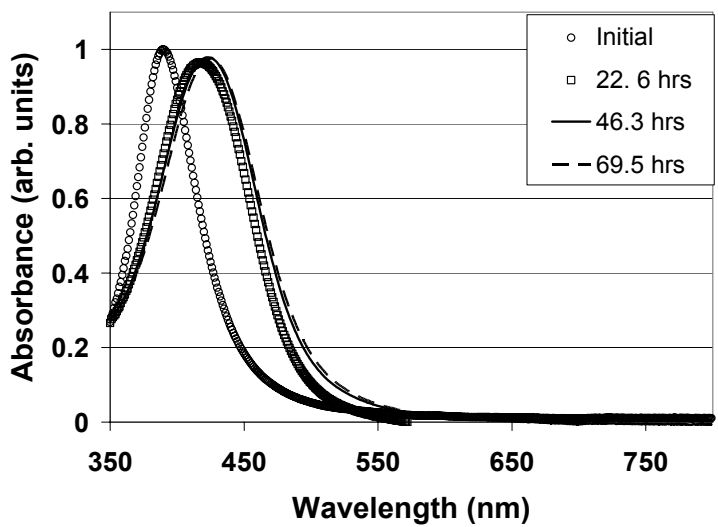

**Figure 8**. Absorption spectrum for the colloid irradiated with a UV lamp.

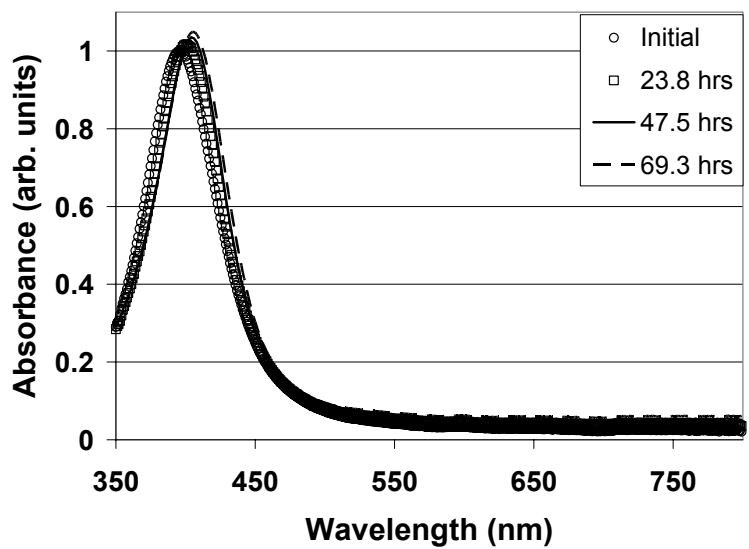

**Figure 9**. Absorption spectrum for the colloid irradiated with a He-Ne laser.

The effects of a fluorescent lamp, an argon-ion laser, a UV lamp, and a He-Ne laser *in the borohydride-based method* are summarized as follows. Irradiation of the silver colloid with light from a fluorescent lamp consisting of multiple lines in the visible range (**Figure 1**) with a power density about 7 mW/cm$^2$ for about 20 hours produced the most nanoprisms. Irradiation with the argon-ion laser ($\lambda$=514.5 nm, 30 mW power) led to the appearance of a second peak in the absorption spectra at 550 nm (**Figure 6**), which grew with longer irradiation. This also caused a shift of the original peak from about 400 nm to 424 nm. After 24 hours the second peak grew to be larger than the original shifted peak. Irradiation by a UV lamp with an output spectrum centered at 365 nm and a much smaller part of the spectrum between 600 and 800 nm and a total power density of about 1.7 mW/cm$^2$ for 24 hours led to a shift of the original absorption peak from 391 nm to 427 nm without a noticeable change in the absorption at wavelengths greater than 550 nm. Longer duration irradiation up to 68 hours did not produce further changes in the absorption spectrum. The helium-neon laser ($\lambda$=632.8 nm) did not produce any noticeable changes in the absorption spectra even after 69 hours of irradiation. When the argon-ion laser or the fluorescent lamp were used, the changes in the absorption spectra were accompanied by a change in the observed color of the solution. The solution irradiated with the argon-ion laser appeared greenish-blue, while the solution irradiated with the fluorescent lamp appeared purple. The solutions irradiated with the helium-neon laser and the UV lamp remained yellow after the irradiation. It appears that irradiation by either the argon-ion laser or the fluorescent lamp typically leads to an increase of the average size of the nanoparticles and to the appearance of randomly-shaped aggregates, while irradiation by the fluorescent lamp leads to more abundant growth of nanoprisms.

*3.2. PVP Method*.

In the second (PVP-based) method, a different, lower-power mercury vapor lamp and an argon-ion laser were used in order to stimulate conversion of silver nanoparticles. In the case of the mercury lamp, a solution of 500.7 mg of PVP in 40mL of distilled deionized water and 160mL of absolute ethanol was heated up to 77

°C in an Erlenmeyer flask equipped with a stir bar. Then 800.5 mg of AgNO$_3$ was added and the solution was heated between 76 and 79 °C continuously for 20 min, while being stirred every 3 minutes. Then the solution was transferred into the appropriate number of quartz cuvettes of 1 x 1 x 4.5 cm size, one of which was irradiated between two mercury lamps at a distance of 10 cm for 25 days. Each lamp had an electrical power of 9 W and an optical power density of about 4 mW/cm$^2$ at a distance of 10 cm. The radiation spectrum of one of these identical mercury lamps is given in **Figure 10**. It shows a substantial part of the spectrum in the visible range between 400 and 700 nm similar to the fluorescent lamp (**Figure 1**).

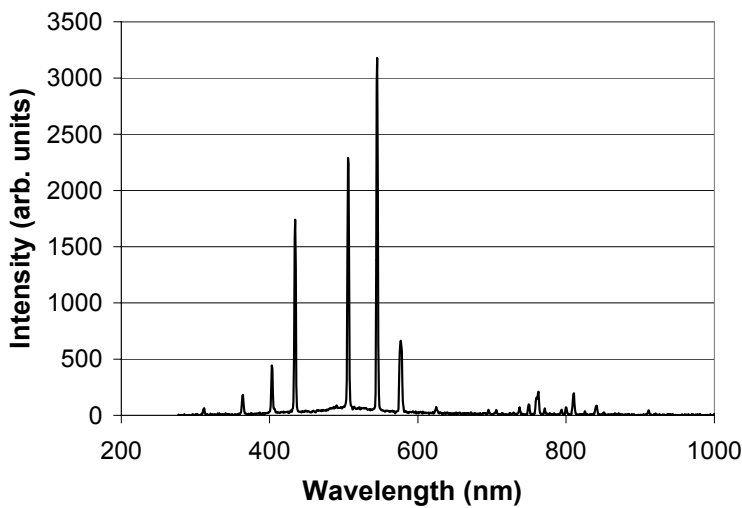

**Figure 10.** Spectrum of the mercury vapor lamp.

With the careful adjustment of the AgNO$_3$/PVP ratio, irradiation caused a red shift of the initial peak and the appearance of an additional red peak in the absorption spectrum, analogous to the changes shown in **Figure 2**. TEM images of nanoparticles from the irradiated samples confirmed that such changes were associated with an increase of the average size of the nanoparticles and with the change of their shape. Examples of these TEM images are shown in **Figures 11** and **12**. Both figures display a mixture of nanoparticles with multifold increased size and with a variety of shapes. Like in the case of irradiation by the fluorescent lamp described above, irradiation of the AgNO$_3$/PVP by the mercury lamps caused the formation of triangular nanoprisms with truncated angles. These are shown at higher resolution in **Figures 13** and **14**.

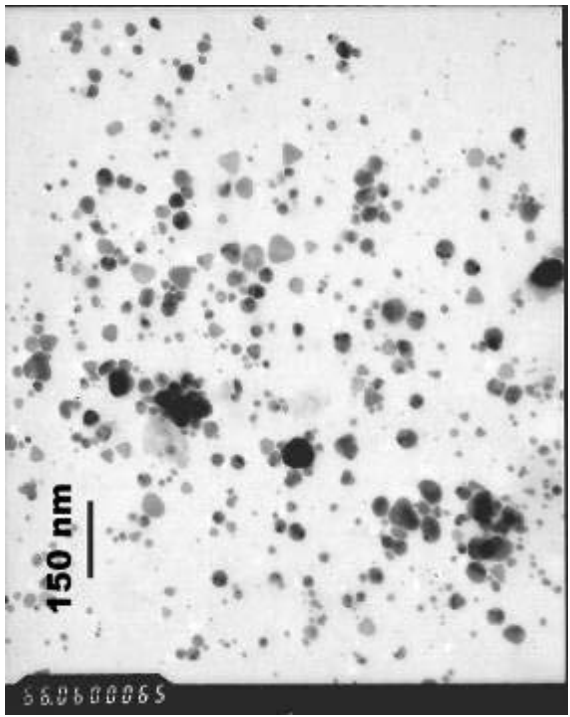

**Figure 11.** A mixture of asymmetric and symmetric silver nanoparticles formed by the photostimulated AgNO$_3$/PVP-based method with irradiation by the mercury vapor lamp.

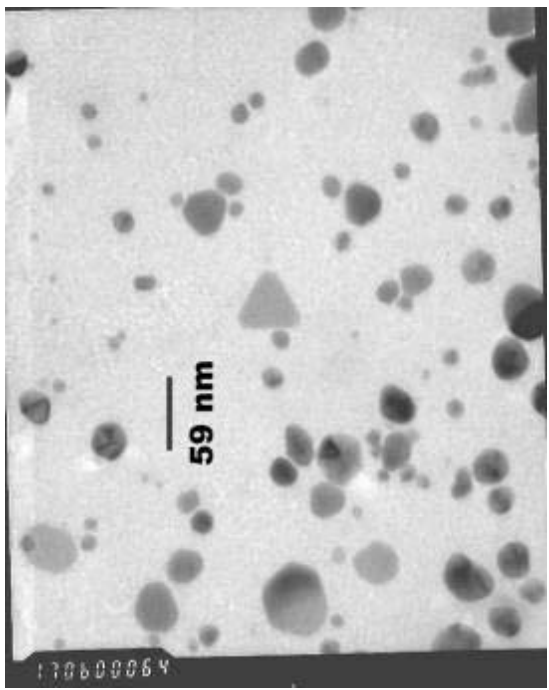

**Figure 12.** A higher-resolution image of a mixture of asymmetric and symmetric silver nanoparticles formed by the photostimulated AgNO$_3$/PVP-based method.

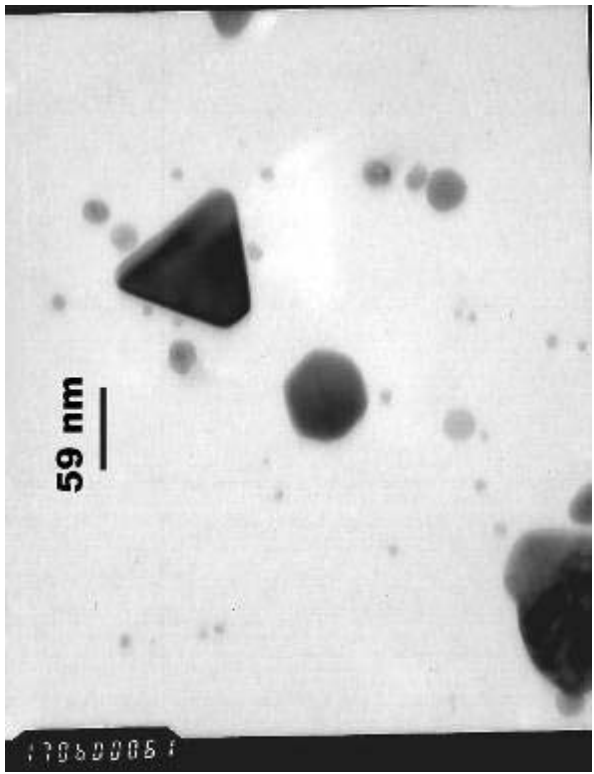

**Figure 13.** Truncated triangular silver nanoprism and nanopolyhedron.

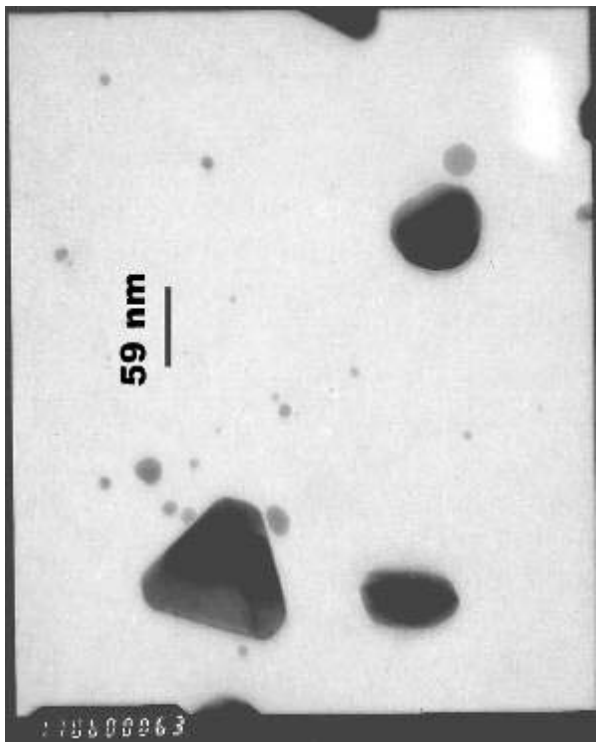

**Figure 14.** Silver nanoprism and nanoellipse.

Our experiments revealed that the abundance of the different types of nanoparticles varied by changing the $AgNO_3$/PVP ratio as well as the irradiation source. More often, nanoprisms were produced with the aid of broadband radiation. Irradiation with an argon-ion laser usually causes the appearance of the additional red-shifted peak in the AgNO$_3$/PVP absorption spectrum, which is likely due to the formation of large nanoparticles of quasi-elliptical shape shown in **Figures 15** and **16** as well as small clusters of large nanoparticles shown in **Figure 17**. In all of the examples described above, the absorption spectra and TEM images of nanoparticles from the colloid samples kept in the dark were essentially unchanged as shown in **Figure 18**. In the course of the experiments, we also found the conditions for the synthesis of huge fractal-type aggregates consisting of hundreds to thousands of small silver nanoparticles. The synthesis was driven by argon-ion laser radiation. The details of this will be reported elsewhere.

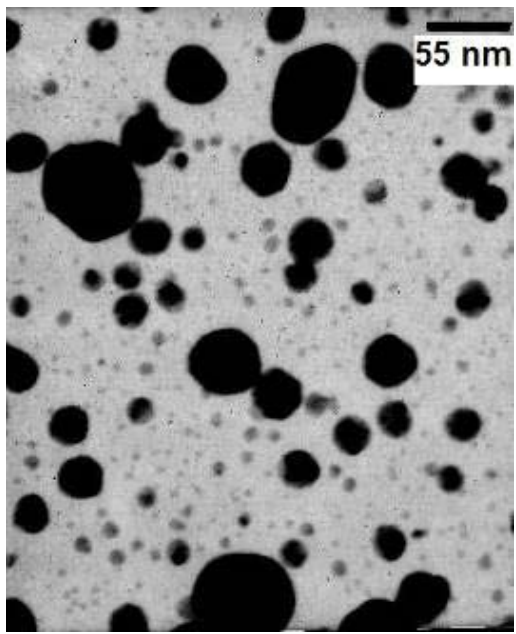

**Figure 15.** Typical silver nanoparticles formed with the PVP-based method and photostimulation with an argon-ion laser.

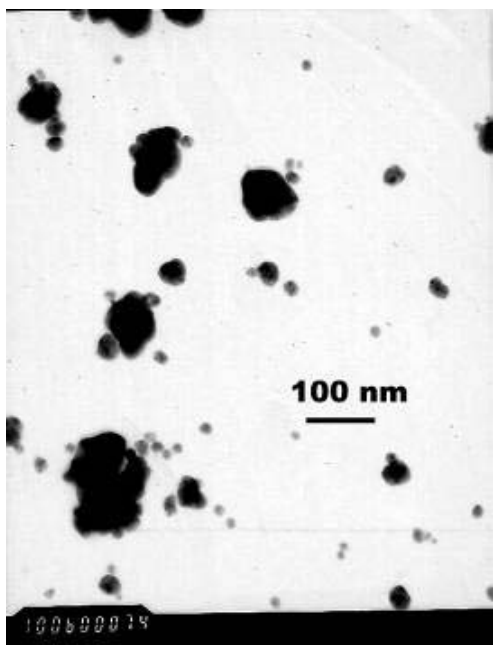

**Figure 16.** Large asymmetric silver nanoparticles formed with the PVP-based method and photostimulation with an argon-ion laser.

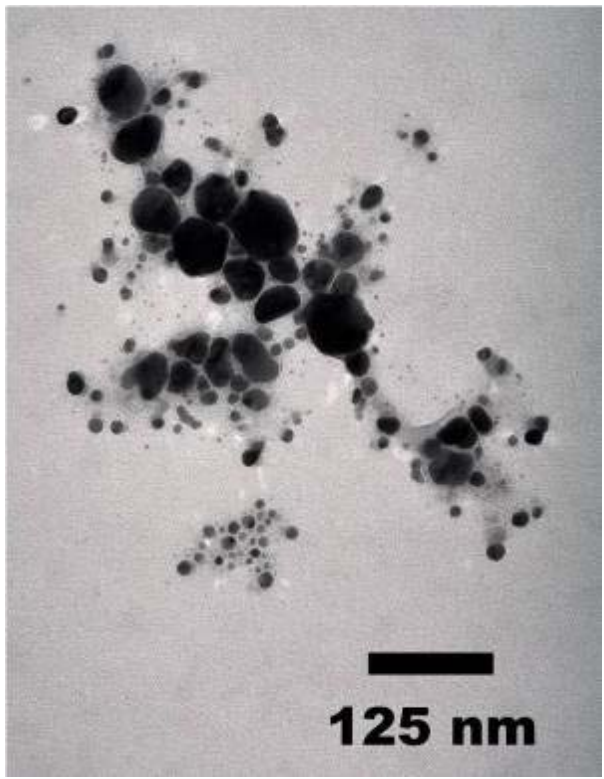

**Figure 17.** A cluster of large, asymmetric, silver nanoparticles formed with the PVP-based method and photostimulation with an argon-ion laser.

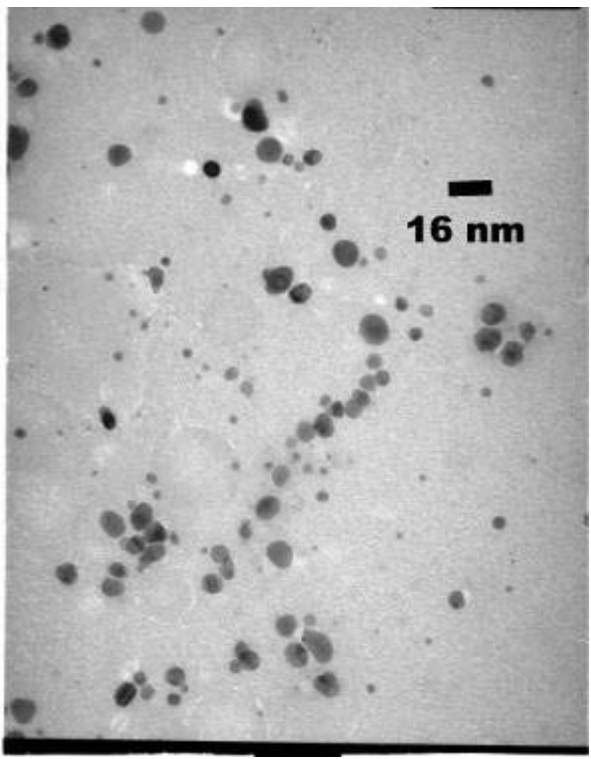

**Figure 18.** A typical TEM image of silver nanoparticles in the colloid kept in the dark.

One concern with sampling should be noted for the TEM images. These images were made by adhering a single drop of the nanoparticle solution onto a carbon film used for the TEM. The single drop used for each sample cannot fully represent the entire solution. Also the use of the electron beam in acquiring the TEM image may somewhat alter the arrangement of the nanoparticles. A final concern regards the repeatability of these experiments. Although the results discussed in this paper have been verified with multiple trials, there were a few trials in which we were unable to observe the typical results. We are trying to determine the reason for these variations.

**Conclusions.**

The specific features of the light-driven synthesis of silver nanoparticles of different sizes and shapes were investigated. The results obtained shed light on the mechanisms behind the manipulation of the morphology of metal nanoparticles driven by the plasmon excitation. They are also important for a variety of nanotechnology-based applications, which require manipulating the shape of metal nanoparticles. The reported experiments proved that the irradiation spectrum, the type of stabilizing agent, and the ratio of the silver-containing and stabilizing agents did appear to have an important affect on the synthesis of the colloidal silver nanostructures. Narrowband red light from a helium-neon laser had no affect on the synthesis. The fluorescent lamp, the argon-ion laser and the mercury lamp caused changes in the absorption spectrum of the colloid and in the sizes and shapes of the produced nanoparticles. Yet, the changes which occurred were different for each light source depending on the method used for the reduction and stabilization of small nanoparticles in the initial, pre-irradiation phase. Changes in the sizes and shapes of isolated nanoparticles as well as the aggregation of small nanoparticles into multiparticle aggregates were observed in both methods using either polychromatic light (400 – 700 nm) or light from an argon-ion laser ($\lambda$=514.5 nm). The borohydride method in combination with the fluorescent lamp seemed to be most suitable for engineering isolated colloidal nanoparticles and manipulation of their sizes and shapes, which includes triangular nanoprisms. This supports the model[5] which suggests the importance of excitation of multiple anisotropic plasmonic resonances associated with nonspherical nanoparticles. Such resonances appear and experience the shift to the red in the course of the particle's growth and shape conversion. Therefore, in the case of broadband visible radiation, such nanoparticles can remain coupled with the stimulating light. Irradiation with the argon-ion laser in combination with the PVP-based method seemed to be most productive for engineering large aggregates consisting of hundreds to thousands of small nanoparticles. The synthesis of large aggregates was found to be very sensitive to the presence of small colloidal aggregates in the initial stage. Most of the nanoprisms observed in the TEM images are approximately the same size, which suggests that they form not necessarily through the merging of smaller ones, but through plasmonic excitation and further conversion of other intermediate asymmetric nanoparticles.

**Acknowledgments.**

A. K. Popov acknowledges the support of this work in part by the grant MDA972-03-1-0020 from DARPA and by the grant 00000496 by ARO. Electron Microscopy support for this project was provided by a University of Wisconsin-Stevens Point Letters and Science Undergraduate Enrichment Initiative grant to R. J. Schmitz.

**References**

1) Shalaev V. M. *Nonlinear Optics of Random Media:* Springer Verlag: Berlin, **2000**. (b) Shalaev V. M. *Optical Properties of Fractal Composites* and *Optical Nonlinearities of Fractal Composites*, Chapters in: *Optical Properties of Random Nanostructures*, Ed: V. M. Shalaev: Springer Verlag, Topics in Applied Physics v.82: Berlin Heidelberg **2002**. (c).Stockman M.I.; Pandey L.N.; Muratov L.S.; George T.F. *Phys. Rev. Lett.* **1994** *72*(15), 2486-2489.

2) Karpov S. V.; Bas'ko A. L.; Popov A. K.; Slabko V. V.; George T. F. *Optics of Nanostructured Fractal Silver Colloids*. Chapter in *Recent Research Developments in Optics 2*, *Part II,* 427-463. Research Signpost: Trivandrum, India, **2002** (see also http://lanl.arxive.org/abs/physics/0301081)

3) Karpov S. V.; Slabko V. V. *Optical and Photo-physical Properties of Fractal-structured Sols of Metals*: Siberian Division of the Russian Academy of Sciences: Novosibirsk, **2003** (in Russian).

4) Kamat P. V. *J. Phys. Chem.B* **2002** 106, 7729-44

5) (a) Jin R. C.; Cao Y.; Mirkin C. A; Kelly K. L.; Schatz G. C.; Zheng J. G. *Science* **2001**, *294*, 1901-03. (b) Jin R. C.; Cao Y.; Hao E.; Metraux G. S.; Schatz G. C.; Mirkin C.A. *Nature* **2003** *425*, 487-490.

6) Noginov M. A.; Vondrova M.; Williams S. M.; Bahoura M.; Gavrilenko V. I., Black S. M.;. Drachev V. P; Shalaev V. M.; Sykes A. *J. Optics A: Pure and Applied Optics* (special issue on Metamaterials) **2005**, *7*, S219-S229. (b) Kim W.; Safonov V. P.; Shalaev V. M.; and Armstrong R. L. *Phys. Rev. Lett*. **1999** *82*, 24-27.

7) Podolskiy V.A.; Sarychev A. K.; Shalaev V. M. *J. Nonlin. Opt. Phys. & Mater*. **2002** 11(1) 65-74. (b) Podolskiy V.A.; Sarychev A. K.; Shalaev V. M. *Optics Express* **2003** *11*, 735. (c) Fang N.; Lee H.; Sun Ch.; Zhang X. *Science* **2005** *308* 534-37.

8) Karpov S. V.; Slabko V. V.; Chiganova G.A. *Colloid Journal* **2002**, *64*, 425-441.

9) (a) Butenko A. V.; Chubakov P. A.; Danilova Yu. E; Karpov S. V.; Popov A. K.; Rautian S. G.; Safonov V. P.; Slabko V. V.; Shalaev V. M.; Stockman M. I. *Zeit. fuer Physik D – At., Mol. Clust.* **1990** *17*, 283-9. (b) Karpov S. V.; Bas'ko A. L.; Popov A. K.; Slabko V. V. *Technical Physics* **2003**, *48*(6), 749-56. (c) Karpov S. V.; Bas'ko A. L.; Popov A. K.; Slabko V. V. *Optics and Spectroscopy*, **2003**, *95(2)*, 230-40. (d) Karpov S. V.; Bas'ko A. L.; Popov A. K.; Slabko V. V. *Optics and Spectroscopy* **2003**, *95(2)*, 241-7.

10) (a) MacDonald K. F.; Fedotov V. A.; Pochon S.; Ross K. J.; Stevens G. C.; Zheludev N. I.; Brocklesby W. S.; Emel'yanov V. I. *Appl. Phys. Lett.* **2002** *80*, 1643-45. (b) Fedotov V. A.; MacDonald K. F.; Zheludev N. I.; Emel'yanov V. I. *J. Appl. Phys.* **2003** *93*, 3540-44.